\documentclass[10pt,a4paper,twocolumn,aps,pra,showpacs]{revtex4-1}
\pdfoutput=1
\usepackage[utf8x]{inputenc}
\usepackage{ucs}
\usepackage{amsmath}
\usepackage{amsfonts}
\usepackage{amssymb}
\usepackage{makeidx}
\usepackage{cellspace,booktabs}
\usepackage{natbib}
\usepackage{lipsum}

\usepackage{color}
\definecolor{light-gray}{gray}{0.55}

\usepackage{microtype}

\usepackage{graphicx}

\renewcommand{\dag}{^{\dagger}}

\newcommand{\bra}[1]{ \langle #1 \rvert }
\newcommand{\ket}[1]{ \lvert #1 \rangle}

\newcommand{\pfrac}[2]{\frac{\partial #1}{\partial #2}}
\newcommand{\intinf}{\int_{-\infty}^{\infty}}

\begin{document}

\begin{abstract}
A Superconducting Quantum Interference Device (SQUID) inserted in a superconducting waveguide resonator imposes current and voltage boundary conditions that makes it suitable as a tuning element for the resonator modes. If such a SQUID element is subject to a periodically varying magnetic flux, the resonator modes acquire frequency side bands.  In this work we calculate the  multi-frequency eigenmodes of resonators coupled to periodically driven SQUIDs and we use the Lagrange formalism to propose a theory for their quantization. The elementary excitations of a multi-frequency mode can couple resonantly to physical systems with different transition frequencies and this makes the resonator an efficient quantum bus for state transfer and coherent quantum operations in hybrid quantum systems. As an example of the application of our multi-frequency modes, we determine their coupling to transmon qubits with different frequencies and we present a bi-chromatic scheme for entanglement and gate operations. 
\end{abstract}

\date{\today}
\author{Christian Kraglund Andersen}
\thanks{E-mail: ctc@phys.au.dk}
\affiliation{Department of Physics and Astronomy, Aarhus University, DK-8000 Aarhus C, Denmark}
\author{Klaus Mølmer}
\affiliation{Department of Physics and Astronomy, Aarhus University, DK-8000 Aarhus C, Denmark}

\title{Multi-frequency modes in superconducting resonators: Bridging frequency gaps in off-resonant couplings}

\pacs{42.50.Pq, 03.67.Lx, 85.25.Dq, 03.67.Bg}

\maketitle

\section{Introduction}
\label{sec:intro}

In the field of circuit Quantum Electrodynamics (circuit QED) the combination of superconducting resonators and Josephson junctions \cite{PhysRevA.69.062320} has been used to demonstrate fundamental quantum interactions. Josphson junctions are also often combined in loops to generate Superconducting Quantum Interference Devices (SQUIDs), which allow more controllability in the systems. The interactions demonstrated in cQED are ranging from the resonant coupling between two-level systems and a harmonic oscillator \cite{Wallraff:2004oh,PhysRevLett.105.060503,PhysRevB.78.180502} over the generation of non-classical states \cite{hofheinz2008generation,vlastakis2013deterministically}, to protocols where the resonator field serves as a quantum bus to transfer quantum states and mediate interactions between different systems \cite{PhysRevLett.107.220501,PhysRevLett.110.250503,RevModPhys.85.623}. To control the interaction between the resonator mode and other physical systems, one uses the ability to tune the frequencies (and sometimes also the damping rates) of the resonator mode \cite{PhysRevLett.107.220501,PhysRevLett.110.250503,RevModPhys.85.623, PhysRevLett.90.238304}. Furthermore, periodic modulation of SQUID parameters is used in parametric amplifiers \cite{flux2008yamamoto} and parametric converters \cite{zakka2011quantum}, where interactions appear between the modes of the resonator \cite{PhysRevA.86.013814}.

\begin{figure}[b]
\flushleft (a)
\includegraphics[width=\linewidth]{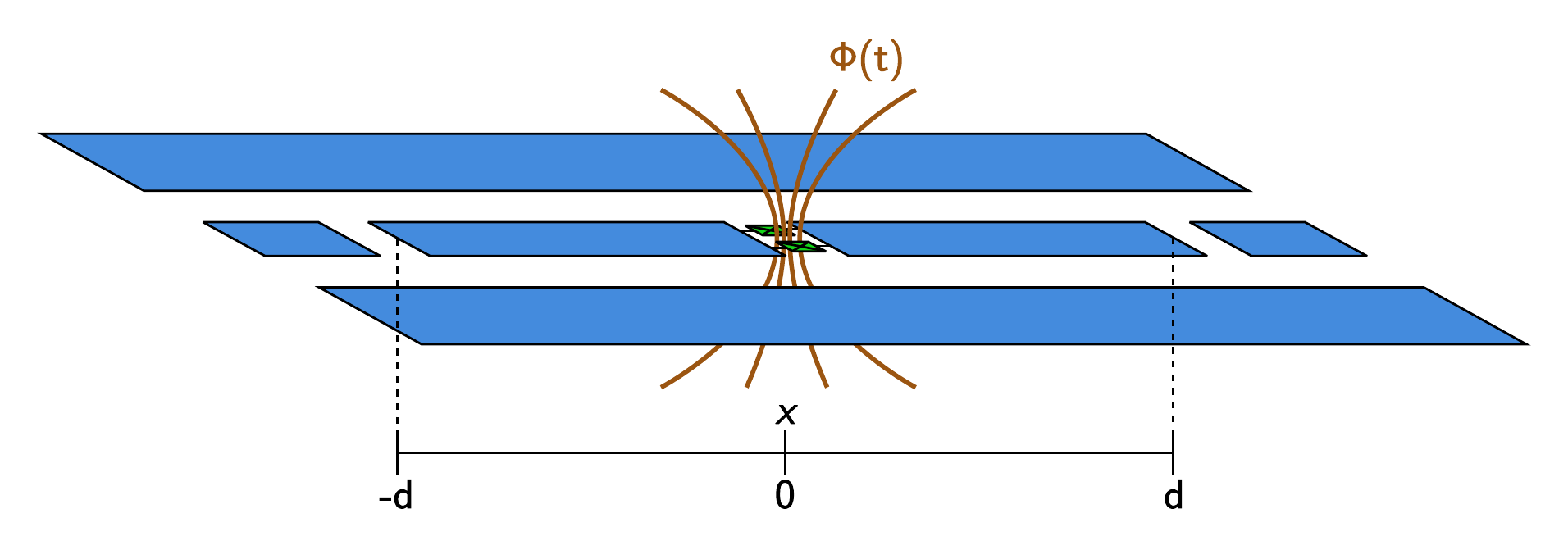}

(b)
\includegraphics[width=\linewidth]{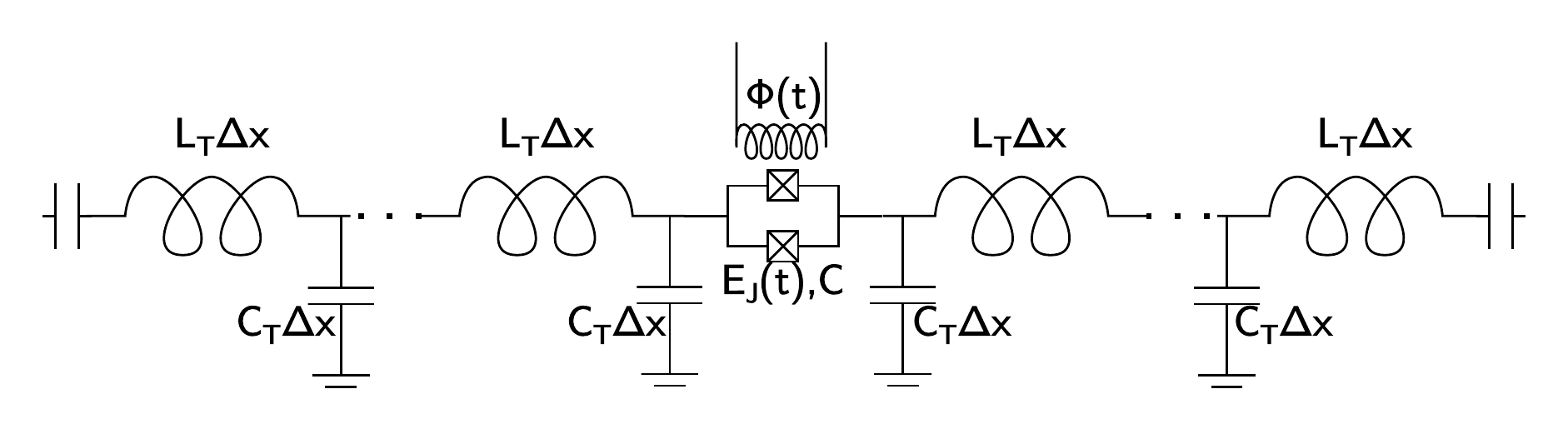}
\caption{(a) A superconducting resonator, interupted by a SQUID at $x=0$. The SQUID is modulated by an external magnetic field $\Phi(t)$ yielding a time dependent Josephson energy $E_J(t)$. (b) Circuit diagram of a resonator interrupted by an inline-SQUID. $L_T$ and $C_T$ denote the inductance and capacitance per length of the waveguide, such that each inductor (capacitor) has the inductance (capacitance) $L_T \Delta x$ ($C_T \Delta x$). The resonator is described in the limit of $\Delta x \rightarrow 0$. The inline-SQUID is characterized by the time-dependent Josephson energy, $E_J(t)$, and its capacitance, $C$.} \label{fig:sketch}
\end{figure}

While changing the frequency of a resonator allows tuning of the photon energies to match energy differences in other systems, the quantized field does not necessarily adiabatically adjust to the change in frequency and the associated change in mode function. Indeed, rapid motion of an optical mirror has been predicted to lead to the creation of photons  by the so-called dynamical Casimir effect \cite{moorequantum}, and experiments in circuit QED have addressed the microwave equivalent by rapid modulation of a SQUID, altering the boundary conditions of a waveguide with half open boundary conditions \cite{PhysRevLett.103.147003,wilson2011observation}. 

In this article we study a similar situation, with a periodically modulated SQUID (see Fig. 1 (a)) in a finite size waveguide with a discrete frequency spectrum. Our goal is not to use the SQUID to drive excitations of the resonator field, and we will hence avoid driving at frequencies supported by the waveguide. Instead we will investigate how the driving leads to new, multi-frequency modes and we will characterize their frequency contents. We then determine how such modes can be used to bridge frequency gaps and coherently couple different physical systems.

Our work is inspired by \cite{flux2008yamamoto, zakka2011quantum, moorequantum, PhysRevA.86.013814, PhysRevLett.103.147003, wilson2011observation} and mechanisms used in atomic memories and light-matter interfaces \cite{RevModPhys.82.1041}, where Raman processes use a laser field to dress atomic or molecular levels such that low frequency transition of the system can be driven and probed at optical frequencies. Similarly, in opto-mechanics \cite{aspelmeyer2010quantum}, weak optical fields impinging on movable mirrors are coherently coupled to low frequency mirror vibrations via the intensity beat node with another field.  By constructing our scheme with circuit QED components, our architecture and the nature of our frequency modulation via changed boundary conditions rather than a non-linear interaction term necessitates a detailed circuit analysis, which we will build on the theory developed in \cite{devoret1995quantum,PhysRevA.86.013814,PhysRevA.89.033853}.

In Sec. \ref{sec:res} we present the physical system and we derive the spatio-temporal solutions to the wave equations for the resonator variables with time dependent boundary conditions. In Sec. \ref{sec:quant}, we present an effective, approximate parametrization of the solutions in terms of canonical conjugate variables, and we establish a quantum theory for the resonator modes and in Sec. \ref{sec:coup} we discuss the interaction across frequency gaps. In Sec. \ref{sec:ms},  we show how our driven modes can establish an effective bi-chromatic entanglement gate operation between two transmon qubits with different transitions frequencies. In Sec. \ref{sec:conc} we present a conclusion and an outlook.

\section{Multi-frequency resonance modes}
\label{sec:res}

To analyze the dynamics of a superconducting resonator modulated by an inline-SQUID we consider the circuit diagram in Fig. \ref{fig:sketch} (b). Here, we have represented the waveguide by a series of $LC$-circuits. Each waveguide segment of length $\Delta x$ is associated with an inductor with inductance $L_T \Delta x$ and a capacitor with capacitance of $C_T \Delta x$. The time-integral of the voltage potential at each node of the circuit constitutes the phase degree of freedom which, in the limit of $\Delta x \rightarrow 0$, becomes a function $\phi(x,t)$ of the continuous position variable $x$ ($L_T$  and $C_T$ are defined as the inductance and capacitance per length of the waveguide). At the center of the waveguide we introduce the inline-SQUID, which contributes a non-linear inductance set by the Josephson energy, $E_J(t)$, and a capacitance set by $C$.

The Lagrangian of the circuit is given by the expression,
\begin{align}
\mathcal{L} =& \int_{-d}^d \bigg\{ \frac{C_T}{2} \big(\partial_t \phi(x,t) \big)^2 - \frac{1}{2L_T} \big( \partial_x \phi(x,t) \big)^2 \bigg\} dx \nonumber\\
&+\frac{C}{2} (\partial_t \Delta \phi(x_J,t))^2+  E_J(t) \cos \frac{2\pi \Delta \phi(x_J,t)}{\Phi_0} , \label{eq:lagrangian}
\end{align}
where $\Delta \phi(x_J,t) = \phi(x_{J+},t) - \phi(x_{J-},t)$ denotes the change in the phase variable, $\phi(x,t)$, imposed by the discrete boundary conditions at the inline-SQUID \cite{PhysRevA.86.013814}. In the following expand the cosine term to second order in $\Delta \phi$. When needed, the higher order terms can be reintroduced, see eg. Eq. \eqref{eq:hnl}, as a pertubation at a later point in the analysis.

The Euler-Lagrange equation,
\begin{align}
\partial_x \pfrac{\mathcal{L}}{(\partial_x (\phi(x,t))} + \partial_t \pfrac{\mathcal{L}}{(\partial_t (\phi(x,t))} - \pfrac{\mathcal{L}}{\phi(x,t)} = 0,
\end{align}
yields the wave-equation for the phase variable along the waveguide
\begin{align}
-v^2 \partial_x^2 \phi(x,t) + \partial_t^2 \phi(x,t) = 0, \label{eq:wavext}
\end{align}
where $v$ denotes the propagation speed of the wave.
We assume that our resonator obeys open boundary conditions at its ends at $x=\pm d$, while at $x_J=0$ the SQUID defines boundary conditions, parametrized by the time-dependent Josephson energy, $E_J(t)$, and the capacitance, $C$, of the SQUID,
\begin{align}
&\pfrac{\phi(x,t)}{x}\Big|_{x=\pm d} = 0 \\
&\frac{1}{L_T} \pfrac{\phi(x,t)}{x}\Big|_{x=0\pm} = C \pfrac{^2 \Delta \phi(0,t)}{t^2} + \frac{(2\pi)^2 E_J(t)}{\Phi_0}  \Delta \phi(0,t). \label{eq:bcxt}
\end{align}
Equations \eqref{eq:wavext}--\eqref{eq:bcxt} define the time-dependent mode functions of the circuit that we shall first calculate and then use as basis for the quantum interaction with additional circuit elements.

In this article we consider a periodically modulated magnetic flux, $\Phi(t)$, through the SQUID leading to a harmonically varying Josephson energy,
\begin{align}
E_J(t) = E_{J,0} + \delta E_J \cos \omega_d t.
\end{align}
If the modulation were not too fast {($\omega_d \ll \omega$), we could have treated the SQUID as a quasi-static component and found the eigenmodes of the system for each value of $E_J(t)$. For not too strong driving ($\delta E_J \ll E_{J,0}$) this would give rise to a variation in the eigenmode frequencies $\omega(t) = \omega'+\delta\omega' \cos \omega_d t$, equivalent to the formation of a central frequency component with sidebands at multiples of the modulation frequency $\omega_d$. A natural \emph{ansatz} for the quadrature operator associated with excitation of the quantized resonator circuit would then be given by
\begin{align}
\hat{q} \approx q_{zpf} (1 + \frac{\delta\omega'}{2\omega'} \cos \omega_d t)(\hat{a}\dag + \hat{a}), \label{eq:qah}
\end{align}
where $\hat{a}$ and $\hat{a}^\dagger$ are the corresponding annihilation and creation operators and $q_{zpf}$ denotes the zero point fluctuations of the corresponding physical observable. While this approach may constitute a good approximation if $\omega_d$ and the modulation amplitude $\delta E_J$ were small, we also expect to find resonances in the system with a central frequency and sideband components when we modulate the SQUID more strongly and with high frequency. A more careful approach is therefore needed to calculate the modes and their frequencies for a larger range of driving parameters and to subsequently quantize the system dynamics.

We thus return to the original, linearized problem with the time-dependent boundary condition and to exploit the periodicity of the driving, we transform our fields into frequency space,
\begin{align}
\phi(x,\omega) = \frac{1}{\sqrt{2\pi}} \intinf \phi(x,t)\, e^{-i\omega t} dt.
\end{align}
We rewrite the wave-equation as
\begin{align}
v^2 \partial_x^2 \phi(x,\omega) + \omega^2 \phi(x,\omega) = 0. \label{eq:wave}
\end{align}
while the boundary conditions in frequency space become:
\begin{align}
\pfrac{\phi(x,\omega)}{x}\Big|_{x=0\pm} &= L_T C \omega^2 \Delta \phi(0,\omega) \nonumber\\&\quad + L_T \frac{(2\pi)^2}{\Phi_0} \frac{E_J(\omega) \otimes \Delta \phi(0,\omega)}{\sqrt{2\pi}}, \label{eq:ftbc}\\
\pfrac{\phi(x,\omega)}{x}\Big|_{x=\pm d} &= 0. \label{eq:openbc}
\end{align}
In Eq. \eqref{eq:ftbc}, $E_J(\omega) \otimes \Delta\phi(0,\omega)$ denotes the convolution product, which is readily calculated since  $E_J(\omega) = E_{J,0} \delta(\omega) + \frac{\delta E_J}{2} (\delta(\omega + \omega_d) + \delta(\omega - \omega_d))$,
\begin{align}
&E_J(\omega) \otimes \Delta\phi(0,\omega) = \nonumber\\&\;\, E_{J,0}\Delta \phi(0,\omega) + \frac{\delta E_J}{2} (\Delta\phi(0,\omega + \omega_d) + \Delta\phi(0,\omega - \omega_d)).
\end{align}
Since the coupling to the SQUID in Eq. \eqref{eq:lagrangian} cancels for even modes ($\Delta\phi(x_J,t)=0$), we consider only odd solutions to Eqs. \eqref{eq:wave} and \eqref{eq:openbc}, i.e., solutions in the form
\begin{align}
\phi(x,\omega) = \begin{cases}\phi(\omega) \cos \big( \omega/v \;(d-x) \big) & \text{for } x > 0, \\ - \phi(\omega)\cos \big( \omega/v \;(-d-x) \big) & \text{for } x < 0. \end{cases}
\end{align}

These functions display the discrete jump at $x=x_J=0$ necessary to fulfil the boundary condition (\ref{eq:ftbc}), which leads to the equation
\begin{align}
 \frac{\omega }{v}d \,\sin \Big(\frac{\omega}{v}d\Big) \phi(\omega) =
2\frac{L_T d}{ L_J}  \cos \Big(\frac{\omega}{v}d\Big)\phi(\omega) \nonumber \\ +\, 2 \frac{L_T d}{\delta L_J} \cos \Big(\frac{\omega + \omega_d}{v}d\Big) \phi(\omega + \omega_d) \nonumber \\ +\, 2 \frac{L_T d}{\delta L_J} \cos \Big(\frac{\omega - \omega_d}{v}d\Big) \phi(\omega - \omega_d), \label{eq:newbc}
\end{align}
 where we have defined $L_J = \sqrt{2\pi} \Phi_0 / (4\pi^2 E_J)$ and $\delta L_J = 2 \sqrt{2\pi} \Phi_0 / (4\pi^2 \delta E_J)$. We have neglected contributions from the Josephson capacitance, assuming typical values obeying $C \omega^2 \ll L_T / L_J$ \cite{PhysRevLett.103.147003}.

Equation \eqref{eq:newbc} can be solved numerically for the allowed discrete values of $\omega$ and the corresponding vector of amplitude strengths $\phi(\omega+m\omega_d)$. For our purpose it is sufficient to approximate the system and restrict ourselves to a dominant frequency component, $\omega$, and two sidebands, $\omega\pm\omega_d$, and ignore coupling to further frequency components $\omega \pm 2 \omega_d, ... $. With this approximation the carrier and sideband components of the system eigenmodes have wave numbers that obey
\begin{align}
kd =& \frac{2L_Td}{L_J} \frac{\cos kd}{\sin kd} + \frac{\big(\frac{2L_Td}{\delta L_J}\big)^2 \frac{\cos kd}{\sin kd}\cos k_-d}{\frac{2L_Td}{ L_J} \cos k_-d + k_-d\sin k_-d} \nonumber \\&+ \frac{\big(\frac{2L_Td}{\delta L_J}\big)^2 \frac{\cos kd}{\sin kd}\cos k_+d}{\frac{2L_Td}{L_J} \cos k_+d + k_+d\sin k_+d}. \label{eq:hard1}
\end{align}
where $k = \omega/v$ and  $k_\pm = (\omega \pm \omega_d)/v$.  The amplitudes of the sidebands are given by
\begin{align}
\phi(\omega \pm \omega_d) = \frac{ \frac{2L_Td}{\delta L_J} \cos k d}{\frac{2L_Td}{L_J} \cos k_\pm d + k_\pm d\sin k_\pm d} \phi(\omega) = A_\pm \phi(\omega). \label{eq:subpm}
\end{align}
We notice that to first order,  the amplitude of the sidebands are inversely proportional to $\delta L_J$ and hence proportional to the driving amplitude $\delta E_J$. This is reminiscent of the important role played by the classical (pump) field amplitude in atom-light interfaces \cite{RevModPhys.82.1041} and in optomechanics \cite{aspelmeyer2010quantum}.

\begin{figure}
\includegraphics[width=\linewidth]{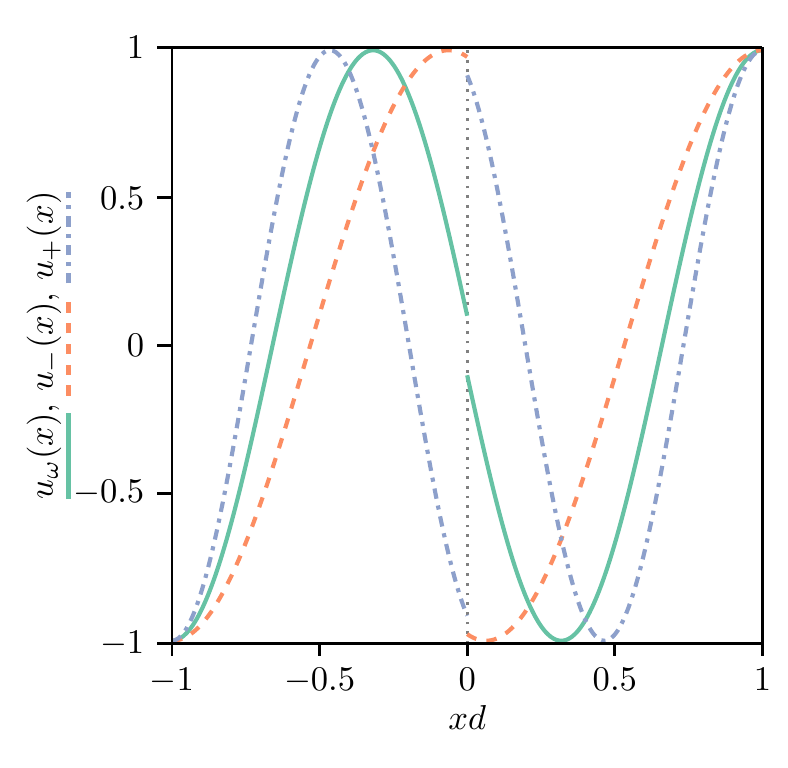}
\vspace*{-1cm}
\caption{The mode-functions when we modulate the SQUID with a frequency $\omega_d = 2\pi \times 2.0$ GHz and $\delta E_J / E_{J,0} = 0.4$ for a mode with $kd = 4.614$ ($\omega = 2\pi \times 7.343$ GHz). The green solid line shows the central frequency component $u_\omega(x)$ while the red dashed and the blue dash-dotted lines show the sideband components $u_-(x)$ and $u_+(x)$ respectively. The parameters of the resonator are $d=1.2$ cm, $E_{J,0}/\hbar = 2\pi \times 715$ GHz, $v = 1.2 \times 10^8$ m/s and $L_T = 50\, \Omega / v$.} \label{fig:mode}
\end{figure}

Now we transform our solution back into the time-domain where the solution to the wave equation attains the form:
\begin{align}
\phi(x,t) &= \text{sign}(x) \cos \big(k(\text{sign}(x)d-x)\big)\cos (\omega t) \phi_{\omega} \nonumber \\
 &\,+ \text{sign}(x) \cos \big(k_+(\text{sign}(x)d-x)\big) \cos(\omega_+  t) \phi_+ \nonumber \\
 &\,+  \text{sign}(x) \cos \big(k_-(\text{sign}(x)d-x)\big) \cos(\omega_-  t) \phi_- \\
 &= u_\omega(x) \phi_\omega(t) + u_+(x) \phi_+(t) + u_-(x) \phi_-(t). \label{eq:three-fr}
\end{align}
In the last line we have separated the terms into space-dependent $u_j(x)$ with values exploring the range $[-1,1]$, while the amplitudes governed by (\ref{eq:subpm}) are included in the time-dependent functions $\phi_j(t) = \cos (\omega_j t) \phi_j$.

We show an example of the mode-structure in Fig. \ref{fig:mode}, where the solid line shows the carrier function $u_\omega(x)$ and the dashed curves show the two associated side band mode functions $u_\pm(x)$. The solutions have three nodes and correspond to the 3rd mode of the resonator. Only odd modes acquire sidebands due to the coupling to the SQUID, and in Fig. \ref{fig:E} we show how the solution of Eq. \eqref{eq:hard1} for the 1st, 3rd and 5th mode leads to eigenfrequencies that vary with the modulation strength, $\delta E_J / E_{J,0}$. In Fig. \ref{fig:E} we notice a frequency shift of the order of 10 MHz, which is much larger than typical bandwidths of superconducting resonators and cannot be predicted without taking the full dynamics into account.

To reach the solution (\ref{eq:three-fr}), we assumed that the sidebands at $\omega \pm \omega_d$ in \eqref{eq:newbc} are present and that the coupling to further components at $\omega\pm 2\omega_d$ is negligible. This assumption is fulfilled for applications in this article, where we deal with $A_{\pm}\sim 10^{-2}$, significantly limiting higher sideband exitation. When the system is externally driven on resonance with the values of $\omega$ shown, or at the sidebands $\omega\pm \omega_d$, the modulation of the SQUID  at $\omega_d$ causes the effective excitation of the multi-frequency solution (\ref{eq:three-fr}). While (\ref{eq:three-fr}) is derived under the assumption of periodic driving, it also holds during transient excitation of the system or coupling to other systems, as long as the resulting evolution is slow with respect to the natural frequencies.

\begin{figure}
\includegraphics[width=\linewidth]{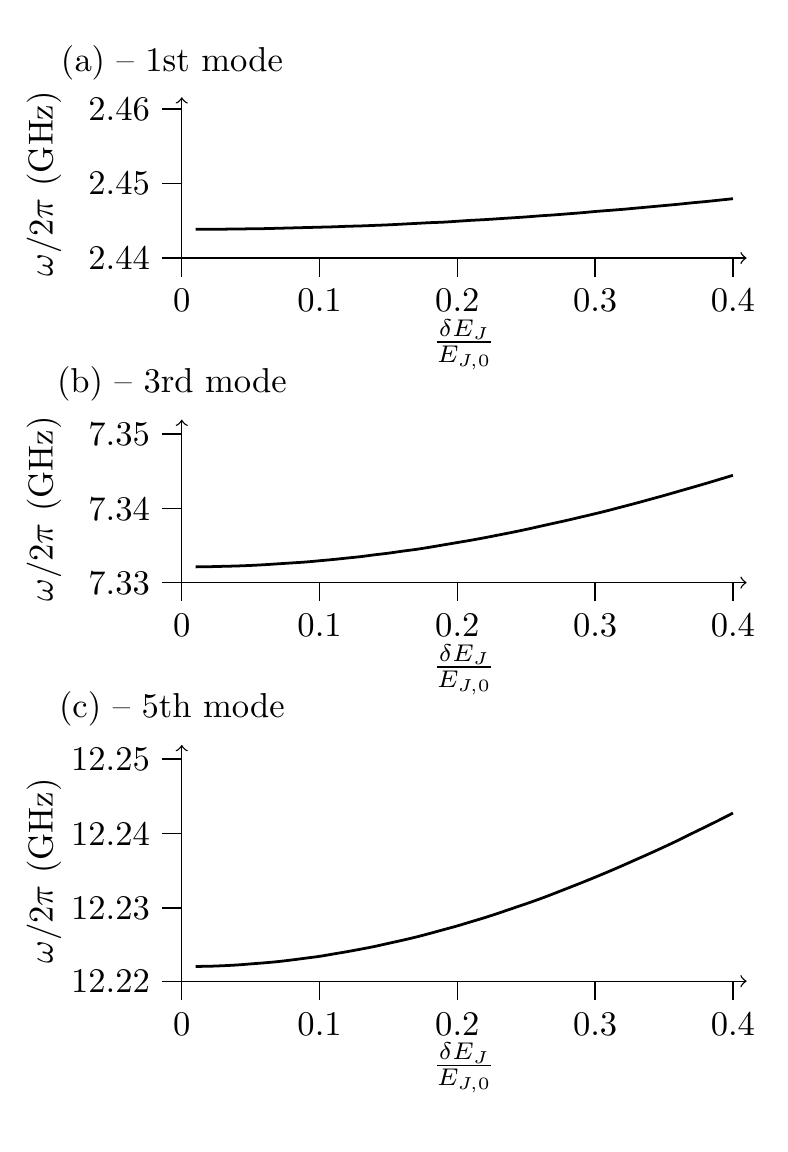}
\vspace*{-1cm}

\caption{Classical resonances for the odd modes of a resonator calculated by Eq. \eqref{eq:hard1} with the same parameters as in Fig. \ref{fig:mode}. We modulate the SQUID with varying strengths at a frequency $\omega_d = 2\pi \times 2.0$ GHz . In (a) we show the variation of the central frequency of the 1st mode, in (b) the 3rd mode and in (c) the 5th mode. The even modes are not considered, since they do not experience the modulation of the SQUID.} \label{fig:E}
\end{figure}

\section{Quantization of the multi-frequency modes}
\label{sec:quant}

Following the normal procedures of second quantization we would have first expanded a general solution $\phi(x,t)$ on stationary eigenmodes of the unmodulated system, and we would have replaced the mode amplitudes by operators satisfying canonical commutation relations. The time-varying SQUID would then be introduced as a driving term which transfers excitation among the quantized modes and which parametrically drives pair-excitation of the modes, cf. \cite{PhysRevA.89.063606, PhysRevLett.113.093602, zhang2014dynamical}. We have deviated from that procedure here because the time varying SQUID modifies the boundary conditions for the resonator modes, and it is clear from Fig. 2. that a large number of the eigenmodes calculated in the absence of the SQUID would be needed to represent the discontinuous jumps in the function $\phi(x,t)$ across the SQUID. Our time dependent mode functions already obey the boundary condition in the presence of the driving. For the finite waveguide, these modes have discrete and well separated frequencies, and we may disregard coupling between them and restrict our analysis to their individual excitation dynamics and their resonant coupling to other systems with transition frequencies at $\omega$ or $\omega\pm \omega_d$. To treat the time dependent mode quantum mechanically, we observe that the central frequency amplitude $\phi_\omega(t)$ oscillates at frequency $\omega$, consistent with the quadratic Lagrangian,
\begin{align}
\mathcal{L} =& \frac{C_\omega}{2} \dot{\phi}_\omega(t)^2 - \frac{1}{2L_\omega} \phi_\omega(t)^2 \label{eq:linlag}
\end{align}
where  $C_\omega$ is conveniently represented by the parts of \eqref{eq:lagrangian} that depend on $\dot{\phi}_\omega(t)$
\begin{align}
C_\omega =&\, 2\int_0^d C_T \cos^2 k (d{-}x) \, dx + C \cos^2 kd \nonumber \\ =&\, C_T d \Big( 1 + \frac{\sin 2kd}{2kd} \Big) + C \cos^2 kd.
\end{align}
The full Lagrangian, indeed, depends on $\dot{\phi}_\omega(t)$ also via the sideband components, but by setting $L_\omega = 1/(C_\omega\omega^2)$ in (\ref{eq:linlag}), we ensure the right evolution frequency of the amplitude variable. The error that we make by assigning the approximate values of $C_\omega$ and $L_\omega$ will only cause relative changes of order $A_{\pm}$ or $A_{\pm}^2$ in the coupling strengths used later in the article.

Introducing the canonical conjugate pair of variables, $\phi = \phi_\omega(t)$, $q = \partial\mathcal{L}/\partial \dot{\phi}_\omega(t) = C_\omega  \dot{\phi}_\omega(t)$, we obtain the effective harmonic oscillator Hamiltonian for the mode
\begin{align}
H = \frac{1}{2C_\omega} q^2 + \frac{1}{2L_\omega} \phi^2.
\end{align}
Applying usual canonical quantization, we impose $[q,\phi]=i\hbar$, and define annihilation and creation operators,
\begin{align}
q = \sqrt{\frac{\hbar C_\omega \omega}{2}} (a\dag + a), \label{eq:qquant} \\
\phi = i\sqrt{\frac{\hbar }{2C_\omega \omega}} (a\dag - a), \label{eq:phiquant}
\end{align}
which obey $[a,a\dag]=1$ and allow rewriting of the quantum mechanical Hamiltonian in the well known form
\begin{align}
H = \hbar \omega (a\dag a + \frac{1}{2}). \label{eq:Hmode}
\end{align}
By construction, this Hamiltonian yields the dynamics of the amplitude of the central frequency component of the driven $\omega$-mode, and Eqs. (\ref{eq:subpm},\ref{eq:three-fr}) account for the physical circuit observables that oscillate at $\omega$ and $\omega_\pm$.

Our  Eqs. \eqref{eq:qquant} and \eqref{eq:phiquant} differ from similar expansions in \cite{PhysRevA.86.013814,PhysRevA.80.032109} where the mode function $u(x)$, and thus the canonical quantum variables, are normalized in terms of the total capacitance $C_\Sigma = 2C_Td + C$ instead of $C_\omega$. Our approach, similar to that of \cite{PhysRevA.89.033853,eichler2014controlling}, gives simpler expressions for our applications.

The validity of the quantization of individual time dependent modes with no mutual couplings relies on the same assumption as we applied to justify the expansion of the classical circuit variables on the multi-frequency eigenmodes. In particular, changes in the excitation of the mode must be slow compared to the frequency separation to other modes.

\subsection{Coupling physical systems across frequency gaps}
\label{sec:coup}

Our quantization of the multi-frequency mode suggests the operator form of the positive and negative frequency parts,
\begin{align}
\hat{\phi}(x,t) = -\frac{i}{2}\sqrt{\frac{\hbar}{2C_{\omega}\omega}} \, a(t)\, \Big(u_\omega(x) + A_+ e^{-i\omega_d t}u_+(x) \nonumber \\ + A_- e^{i\omega_d t}u_-(x) \Big) + \text{h.c.}, \label{eq:qmode}
\end{align}
where $a(t)$ is given in the Heisenberg picture, and therefore $\hat{\phi}(x,t)$, due to (\ref{eq:Hmode}), acquires oscillations at the frequencies $\omega$ and $\omega\pm\omega_d$. From $\hat{\phi}(x,t)$ we can express all other variable of interest like the voltage, charge distribution etc. As an example we can consider the voltage operator 
\begin{align}
\hat{V}(x,t) &= \pfrac{\hat{\phi}(x,t)}{t} \\&= \frac{1}{2}\sqrt{\frac{\hbar}{2C_{\omega}\omega}} \, a(t)\, \Big( \omega u_\omega(x) + \omega_+ A_+ e^{-i\omega_d t}u_+(x) \nonumber \\ &\quad + \omega_- A_- e^{i\omega_d t}u_-(x) \Big) + \text{h.c.}.\label{eq:voltage}
\end{align}

Electric charges and dipoles couple to the voltage along the waveguide, and the expression (\ref{eq:voltage}) implies that the driving of the in-line SQUID allows to bridge the frequency gap between the quantized circuit degrees of freedom and auxiliary quantum systems if they have transition frequencies equal to any one of the frequencies $\omega$ and $\omega\pm\omega_d$  - in the same way as a pump laser field may assist a quantized optical probe in the driving of atomic Raman transitions and optomechanical motion.

The coupling may have different forms, but if a quantum system $\mathcal{S}$ is detuned by a small amount $\delta$, from one of the multimode frequency components, the joint system dynamics is given by a Hamiltonian, which in the interaction picture takes the form,
\begin{align}
H = \hbar \delta\, a\dag a + \hbar G (a\dag b + b\dag a), \label{eq:HI}
\end{align}
where the coupling strength $G$ depends on the physical coupling mechanism and $b$ ($b\dag$) is the lowering (raising) operator of excitations in $\mathcal{S}$. This Hamiltonian is known as a beam-splitter interaction, which adiabatically transfers the quantum state of $\mathcal{S}$ to the resonator. For the application of the resonator as a quantum bus this is the desired interaction. For a physical component with excitation frequency $\omega_S \sim \omega \pm \omega_d$, situated at $x=x_t$ and coupled to the local value of $\hat{V}(x,t)$, we obtain $G_{\pm} \propto (\omega\pm \omega_d) A_{\pm}u_{\pm}(x_t)$, while the value $G \propto \omega u_\omega(x_t)$ is obtained when $\omega_S \sim \omega$. We will write this coupling for the former case as
\begin{align}
G_{\pm} = \frac{g\,\omega_\pm A_\pm}{4 \sqrt{\hbar C_\omega \omega}} u_\pm (x_t), \label{eq:G}
\end{align}
where $g$ is determined by the auxiliary system observables.

As a simple example we consider a transmon capacitively coupled to the resonator \cite{PhysRevA.76.042319} at position $x_t$. The transmon has quantized charge and phase variables $\hat{n}$ and $\hat{\psi}$, such that its Hamiltonian is given by $H_t = 4E_C (\hat{n} - n_g)^2 - E_{J,t} \cos \hat\psi$, while it couples to the resonator with
\begin{align}
H_I = 2e \beta \hat{n}\, \hat{V}(x_t)
\end{align}
where $\beta = C_c / C_\Sigma$ denotes the ratio between the coupling capacitor and the total capacitance of the transmon. In the limit of $E_{J,t} \ll E_C$, the transmon can be approximated as a two-level system, and the charge operator is given by Pauli transition operators between its eigenstates
\begin{align}
\hat{n} = i\Big(\frac{E_{J,t}}{8E_C} \Big)^{1/4} \frac{1}{\sqrt{2}} (\sigma\dag - \sigma).
\end{align}
The coupling constant $g$ between the system states in (\ref{eq:G}) thus becomes \cite{PhysRevA.76.042319}
\begin{align}
g = 2e \beta \bra{1}\hat{n}\ket{0}= \sqrt{2}ie\beta\Big(\frac{E_{J,t}}{8E_C} \Big)^{1/4}.
\end{align}
For realistic transmon and resonator parameters, the coupling strength $G_-$ to the lower sideband of the first mode of a short resonator is shown in Fig. \ref{fig:G} for different values of $\omega_d$. Characteristic resonant coupling strengths of transmon qubits to coplanar waveguides are a few hundred MHz \cite{PhysRevB.77.180502}, and we expect, within the validity of our approximations, to obtain few MHz coupling strengths between transmons and waveguides when an GHz frequency separation between them is bridged by the modulated SQUID. We, indeed, observe such values in the figure, and we also observe the expected increase of the coupling strength with the modulation amplitude $\delta E_J$. Due to the factor of $\omega_-$ in the expression for $G$ we obtain a lower value when we modulate with a higher frequency. In the figure we have also included a calculation using the simplified approach presented in the beginning of Sec. \ref{sec:res}. This quasi-static approach yields a coupling that is independent of the modulation frequency and in the regime of small modulation frequency and amplitude it matches our more precise calculation.

 We have chosen parameters to ensure a large free spectral range for the transmon qubits, since close to a resonance a cross-Kerr effect between the transmon and the resonator mode of strength
\begin{align}
\chi = G_\omega^2 \frac{\alpha}{\hbar \Delta (\hbar \Delta + \alpha)}, \label{eq:kerr}
\end{align}
may alter the dynamics. In Eq.(\ref{eq:kerr}), $\Delta = \omega_{10} - \omega$, $G_\omega$ is the coupling strength to the center frequency and $\alpha = -E_C$ is the anharmonicity of the transmon \cite{PhysRevA.76.042319}. For the parameters used for the thick green line in Fig. \ref{fig:G} $\chi \approx 2\pi \times 0.2$ MHz, and it is independent of the modulation frequency amplitude.

Since the same mode may be simultaneously coupled to different systems via its different frequency components, it may serve to transfer quantum states coherently between systems whose frequencies differ by $\omega_d$ or $2\omega_d$. Since we may apply a more complicated signal to the inline-SQUID, we may also modulate it at different frequencies and thus establish sidebands, which can interact selectively with different systems.  In the next section we shall couple the multi-frequency mode to two different transmon qubits and obtain a bichromatic two-qubit entangling gate, similar to gates applied in ion traps.

\begin{figure}
\includegraphics[width=\linewidth]{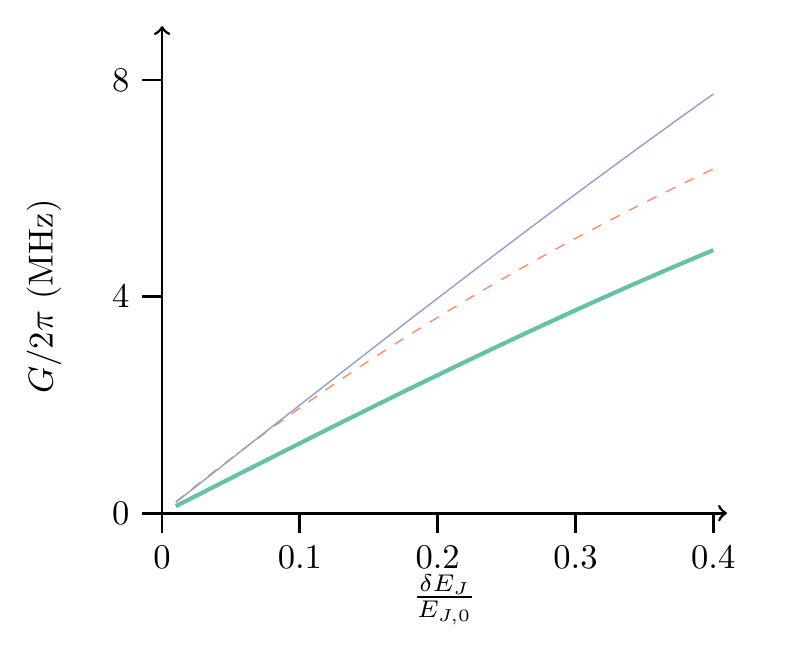}
\vspace*{-1cm}

\caption{Coupling strength of the two lowest states of a transmon to the resonator mode. The frequencies are such that $\omega_{10} = \omega_- = \omega - \omega_d$, where $\omega$ is the frequency of the 1st mode in a resonator with $d=0.25$ cm and $\omega=2\pi \times 10.82$ GHz in the absence of modulation. The transmon is located at $x_t=0.1d$ and it has $E_J / E_C = 80$ and $\beta = 2/3$. The thick (green) line is calculated by Eq. \eqref{eq:G} for the modulation frequency set to $\omega_d = 2\pi\times 6$ GHz while the thin (blue) line shows  Eq. \eqref{eq:G} for $\omega_d = 2\pi\times 0.5$ GHz. The dashed (red) line shows the result of the quasi-static aspproximation, Eq. \eqref{eq:qah}, which is independent of the modulation frequency.} \label{fig:G}
\end{figure}

\section{Multi-frequency modulation and Multi-Qubit gates}
\label{sec:ms}

\begin{figure}[t]
\flushleft (a)
\includegraphics[width=\linewidth]{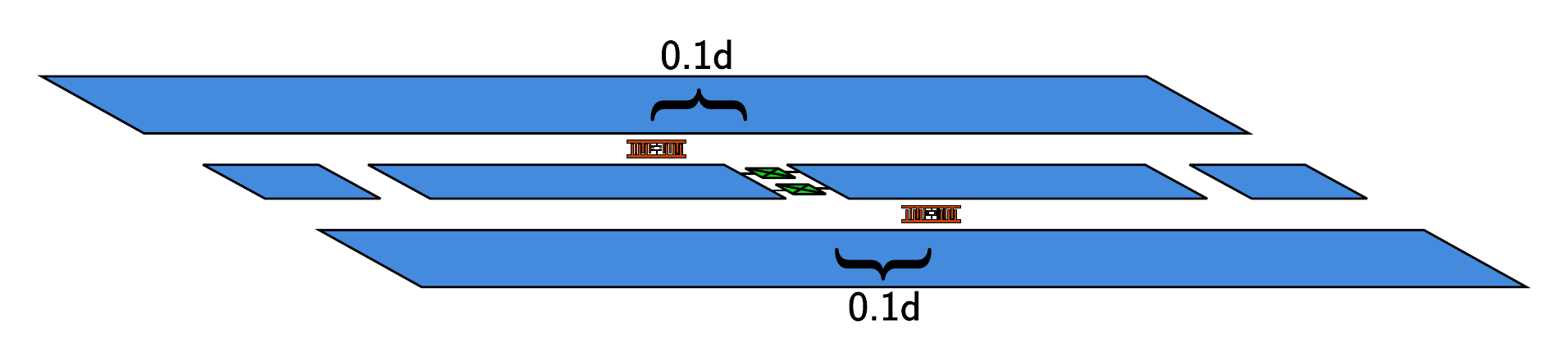}

(b)
\includegraphics[width=\linewidth]{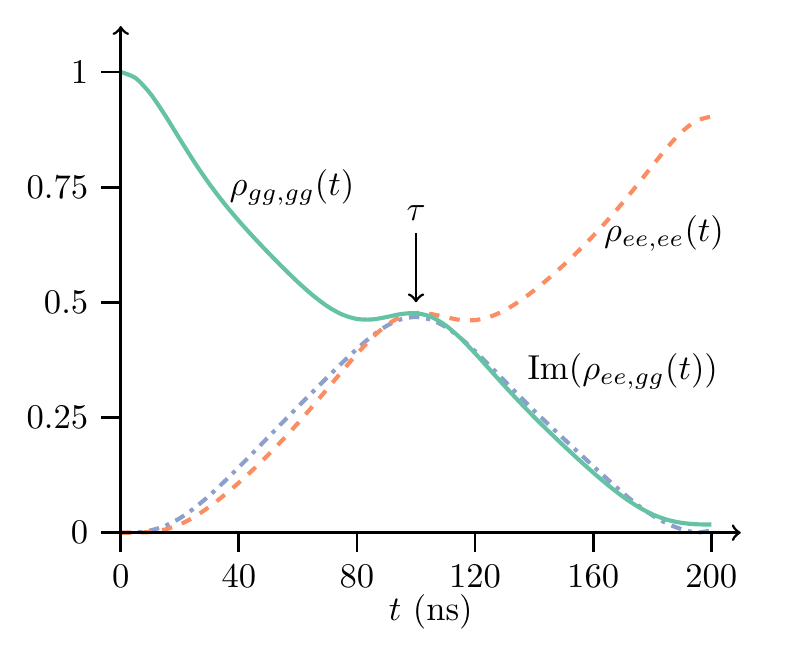}
\caption{(a) Setup with two transmon qubits placed at $x_I=0.1d$ and $x_I=-0.1d$. (b) Time evolution of two-qubit density matrix elements obtained by a simulation of the master-equation with the Hamiltonian in Eq. \eqref{eq:ms}, a resonator decay-rate of $\kappa = 2\pi \times 0.2$ MHz, and transmon decay and dephasing lifetimes $T_1 = 10$ $\mu$s and $T_2 = 5$ $\mu$s. The transmon qubits have different excitation frequencies $\Omega_{1} = 2\pi \times 6.0$ GHz and $\Omega_2 = 2\pi \times 6.5$ GHz and the resonator frequency is $2\pi\times10.82$GHz. For parameters described in the text we obtain the Hamiltonian \eqref{eq:ms} with $G = 2\pi \times 2.5$ MHz and $\delta = 2\pi \times 10$ MHz. In the simulation we have also included a kerr-term with magnitude given in Eq. \eqref{eq:kerr}.} \label{fig:ms}
\end{figure}

Trapped ions can be excited by laser fields at a frequency sideband that excites their collective vibrational motion, and a bichromatic scheme, using laser frequencies detuned symmetrically around the internal state transition frequency, can be used to entangle the internal state of two or more ions \cite{PhysRevLett.82.1971,sackett2000experimental,PhysRevA.62.022311}. In this section we will develop a scheme that uses the multi-frequency resonator modes to accomplish a similar entangling operation among transmon qubits. If the central frequency component of the quantized field has frequency $\omega$, and the qubits have frequency $\Omega$, driving the SQUID at $\omega_t=\omega-\Omega$  makes the transfer of excitation between the qubits and the lower multi-frequency sideband  almost resonant. If, at the same time, the SQUID is driven at
$\omega_p=\omega+\Omega$, a parametric interaction, leading to the joint excitation (and deexcitation) of the resonator mode and the transmons becomes resonant. Detuning of these two driving terms by a small amount $\delta$ leads in the interaction picture with respect to the transmon and single mode Hamiltonian to the near resonant coupling of the transmons and resonator field operators,
\begin{align}
H_I =& \,\hbar G \sum_{n=1,2} (  a e^{-i\delta t} + a\dag e^{i\delta t} )( \sigma_{+,n} + \sigma_{-,n}). \label{eq:ms}
\end{align}
Unlike the ion trap implementation, where laser frequencies differing by a relatively small amount are applied to the ion qubits, our driving fields are applied to the resonator system and their frequencies differ strongly (by twice the qubit transition frequency, $\omega_p-\omega_t \sim 2\Omega$). The interaction Hamiltonian, however, is the same, and the analysis in \cite{PhysRevA.62.022311} applies for both the trapped ions and for the superconducting qubit system. We can hence use the scheme  to accomplish a two-qubit entangling gate as part of a universal gate set and with more transmon qubits, we may also prepare multi-qubit entangled states.

Before passing to an example with realistic interaction and damping parameters, we note that transmon qubits are likely to have different transition frequencies. If these are well separated, we merely have to modulate the in-line SQUID by separate pairs of modulation frequencies, $\omega_{t,n}=\omega-\delta-\Omega_n$, $\omega_{p,n}=\omega-\delta+\Omega_n$ and strengths, in which case we can recover Eq. (\ref{eq:ms}). In Fig. 5 we illustrate a coplanar waveguide resonator coupled to two transmon qubits with different transition frequencies, $\Omega_{1} = 2\pi \times 6.0$ GHz and $\Omega_2 = 2\pi \times 6.5$ GHz. We modulate the SQUID at frequencies $\lbrace \omega_{t,1}, \omega_{p,1}, \omega_{t,2}, \omega_{p,2} \rbrace = 2\pi \times  \lbrace 4.366, 17.366, 4.866, 16.866 \rbrace$ GHz and with the amplitudes $\lbrace \delta E_{J,t1}, \delta E_{J,p1}, \delta E_{J,t2}, \delta E_{J,p2} \rbrace / E_{J,0} = \lbrace 0.1581, -0.1584, 0.1682, -0.1687\rbrace$, leading to the Hamiltonian (\ref{eq:ms}), with $G = 2\pi \times 2.5$ MHz and $\delta = 2\pi \times 10$ MHz. With these parameters, an entangling gate has the duration $\tau=2\pi/\delta = 100$  ns \cite{PhysRevA.62.022311}, as indicated in Fig. 5 (b) by the time evolution of the two-qubit density matrix elements. The three matrix elements shown in Fig. \ref{fig:ms} are, in the ideal case, the only non-zero matrix elements of the two-qubit density matrix and thus the concurrence can be expressed as $2\, \text{Im}(\rho_{ee,gg})$. In the calculation, we have taken the finite coherence and excitation lifetimes of the qubit and of the resonator into account, and we obtain a maximally entangled state with a fidelity $F=95 \%$.

Our proposal is competitive with the achievements of other entanglement-gates in circuit QED \cite{dicarlo2009demonstration, PhysRevLett.107.080502, PhysRevLett.109.060501, PhysRevA.88.032317, PhysRevA.87.022309} and it may offers some advantages: (i) Using the SQUID to modulate the resonator frequency, there is no need for additional control lines to the qubits and we avoid the dephasing associated with the conventional frequency tuning of qubits. (ii) When our qubits are idle, they are far detuned with respect to each other and to the resonator, and hence they are immune to cross talk and Purcell-enhanced decay.

\section{Conclusion and outlook}
\label{sec:conc}

In conclusion, we have identified the resonant modes of a superconducting resonator with periodically modulated boundary conditions. The resulting modes consist of a carrier and two weak sideband components at frequencies which are separated by the modulation frequency, see Eq. \eqref{eq:three-fr} and Fig. \ref{fig:mode}. The classical sideband amplitudes are proportional to the carrier amplitude, and assuming that this proportionality is maintained, the dynamics is described by a simple harmonic oscillator Hamiltonian, which we take as the starting point for our quantum analysis. Elementary excitations of the system thus contain three frequency components and can exploit the sidebands to couple different physical system separated by large frequency gaps. Our sideband mediated transfer of excitation between two systems with very different frequencies can also be viewed as a transition between two energy eigenstates of the joint system, driven resonantly by the oscillating magnetic flux applied to the inline-SQUID. The rapid modification of SQUID parameters is already routinely accomplished for rapid tuning in laboratories and the harmonic driving is used in parametric amplifiers \cite{flux2008yamamoto} and parametric converters \cite{zakka2011quantum}. We hence believe that our proposal can be implemented with devices that can be readily constructed, while modulation of tuning elements to form sidebands rather than static frequency shifts may be employed in resonators that are already in use. 

The multi-frequency modes were calculated using a linearized Lagrangian. Treating the non-linearity of the Josephson Hamiltonian as a perturbation on the single mode Hamiltonian, we obtain the Kerr-nonlinearity \cite{PhysRevA.86.013814,eichler2014controlling},
\begin{align}
H_{NL} = -\frac{E_{J,0}}{4} \Bigg( \frac{2\pi \sqrt{\hbar/2C_\omega \omega}}{\Phi_0} \Bigg) \cos^4 kd\; a\dag a\dag a a. \label{eq:hnl}
\end{align}
This term is small for the parameters used  in this paper, but in combination with the ability to couple systems at very different frequencies, its application, e.g., for the generation of non-classical states of the oscillator mode may be useful.

Finally, we discussed the application of the quantized multi-frequency oscillator mode to mediate an entangling gate between two transmon qubits, and by simulation we showed that a Bell-state with a fidelity of 95\% may be achieved using realistic parameters. This  analysis readily generalizes to multi-qubit systems, where the frequency control of the SQUID modulation may be used to control one and two-qubit gates as well as multi-qubit entanglement operations on any qubits in the resonator. 

A bulk acoustic wave-modulator has been recently proposed to modulate the capacitance of a superconducting $LC$-circuit \cite{PhysRevLett.108.130504} and hence mediate the frequency difference between the quantized high frequency excitations of the circuit and the slow motion of a trapped atomic ion. That proposal relies in a similar way as ours on the frequency modulation of a quantized field degree of freedom, and due to the modulated capacitance it couples more strongly to low frequency motion than our oscillating voltage (\ref{eq:voltage}), which carries prefactors proportional to the low frequency. 

\section*{Acknowledgements}
We thank P. Haikka, A.C.J. Wade, M.C. Tichy and Q. Xu for feedback on the manuscript. We furthermore acknowledge support from the Villum Foundation Center of Excellence, QUSCOPE, and the EU 7th Framework Programme collaborative project iQIT.

\bibliography{bt}

\end{document}